\documentclass[11pt]{article}

\usepackage[utf8]{inputenc}
\usepackage[T1]{fontenc}
\usepackage[margin=1in]{geometry}

\usepackage{amsmath,amssymb,amsthm,mathtools}
\usepackage{array}
\usepackage{booktabs}
\usepackage{graphicx}
\usepackage{tabularx}
\usepackage{xcolor}
\usepackage{enumitem}
\usepackage{setspace}
\usepackage{hyperref}
\usepackage{xurl}
\usepackage[round,authoryear]{natbib}
\usepackage{etoolbox}
\usepackage{titling}
\usepackage{tcolorbox}

\tcbset{
  colback=black!5,
  colframe=black!50,
  boxrule=0.5pt,
  arc=1mm,
  left=2mm,right=2mm,top=2mm,bottom=2mm
}

\hypersetup{
  colorlinks=true,
  linkcolor=blue!50!black,
  citecolor=blue!50!black,
  urlcolor=blue!50!black,
  bookmarksopen=true,
  bookmarksnumbered=true
}

\newcolumntype{Y}{>{\raggedright\arraybackslash}X}
\newcommand{\layerbox}[2]{%
  \fbox{%
    \parbox{0.92\linewidth}{%
      \textbf{#1}\\[-0.2em]
      #2%
    }%
  }%
}

\theoremstyle{plain}
\newtheorem{proposition}{Proposition}
\newtheorem{hypothesis}{Hypothesis}

\theoremstyle{definition}
\newtheorem{definition}{Definition}

\title{\textbf{AI Agents in Financial Markets: Architecture, Applications, and Systemic Implications}}
\author{Hui Gong\texorpdfstring{\protect\footnotemark[2]}{}}
\date{April 2026}

\begin{document}

\begingroup
\renewcommand{\thefootnote}{\fnsymbol{footnote}}
\maketitle
\footnotetext[2]{UCL Institute of Finance \& Technology, University College London. Email: \texttt{h.gong.12@ucl.ac.uk}.}
\endgroup
\setcounter{footnote}{0}

\vspace{-0.8em}

\begin{abstract}
Recent advances in large language models, tool-using agents, and financial machine learning are shifting financial automation from isolated prediction tasks to integrated decision systems that can perceive information, reason over objectives, and generate or execute actions. The paper develops an integrative framework for analysing agentic finance: financial market environments in which autonomous or semi-autonomous AI systems participate in information processing, decision support, monitoring, and execution workflows. The analysis proceeds in three steps. First, the paper proposes a four-layer architecture of financial AI agents covering data perception, reasoning engines, strategy generation, and execution with control. Second, it introduces the Agentic Financial Market Model (AFMM), a stylised agent-based representation linking agent design parameters such as autonomy depth, heterogeneity, execution coupling, infrastructure concentration, and supervisory observability to market-level outcomes including efficiency, liquidity resilience, volatility, and systemic risk. Third, it presents an illustrative empirical application based on event studies of AI-agent capability disclosures and heterogeneous market repricing. It argues that the systemic implications of AI in finance depend less on model intelligence alone than on how agent architectures are distributed, coupled, and governed across institutions. The empirical application is intentionally exploratory: it does not validate the full AFMM, but shows how one observable expectations channel can be studied using public data. In the near term, the most plausible equilibrium is bounded autonomy, in which AI agents operate as supervised co-pilots, monitoring systems, and constrained execution modules embedded within human decision processes.
\end{abstract}

\vspace{0.4em}


\vspace{0.25em}


\vspace{1em}

\section{Introduction}\label{sec:introduction}

Financial markets have always rewarded speed, scale, and quality in information processing. For decades, advances in market infrastructure and quantitative modelling have enabled progressively faster execution, richer data analysis, and more automated decision pipelines. What is changing now is not simply that artificial intelligence (AI) is becoming more accurate at prediction, but that it is beginning to operate across the \emph{full workflow} of financial decision-making. Recent developments in large language models (LLMs), retrieval-augmented generation, tool-using agents, memory systems, and autonomous planning modules have made it increasingly feasible to build systems that can interpret heterogeneous data, reason over objectives, generate action proposals, and interact with external tools in sequential environments \citep{yao2023react, park2023generative, xi2023rise}.

This shift has particular significance in finance. Financial decisions are rarely one-shot prediction problems. They typically require the integration of structured market data, unstructured textual information, institutional constraints, risk limits, and execution rules. A portfolio manager, trader, compliance officer, or supervisor does not merely classify an input; they must move through a sequence of steps that includes information gathering, interpretation, prioritisation, action selection, and post-decision monitoring. In this respect, the emergence of AI agents marks a potentially important transition from \emph{model-centric automation} to \emph{workflow-centric automation}. Rather than assisting with only one component of the chain, agentic systems may increasingly participate in the end-to-end process through which financial actions are formulated, evaluated, and, in some settings, executed.

The timing of this transition is not accidental. Over the past few years, the technical foundations for agentic finance have advanced rapidly. On the language-model side, finance-specific systems such as BloombergGPT and FinGPT have demonstrated the growing viability of domain-adapted LLMs for financial text understanding and analysis \citep{wu2023bloomberggpt, yang2023fingpt}. Benchmark studies such as FinBen further suggest that financial competence cannot be reduced to generic language ability alone, and that finance-specific reasoning remains an important frontier \citep{xie2024finben}. At the same time, more recent work has moved beyond static NLP tasks toward reasoning-oriented and agentic designs, including systems that combine planning, memory, tool use, and multi-agent coordination for financially relevant tasks \citep{liu2025finr1, wu2025finteam, xiao2024tradingagents}. These developments suggest that the relevant research question is no longer whether AI will be used in finance, but how increasingly agentic forms of AI will reshape financial workflows and market structure.

The policy context reinforces the urgency of this question. Recent analyses by the OECD, the Financial Stability Board (FSB), and the Bank for International Settlements (BIS) all point to a dual dynamic: on the one hand, AI can improve efficiency, monitoring, compliance, and analytical capacity; on the other hand, it may amplify existing vulnerabilities through third-party dependencies, correlated model behaviour, governance weaknesses, cyber risks, and the rapid propagation of machine-mediated decisions \citep{oecd2023genaifinance, oecd2024regulatory, fsb2024ai, aquilina2025harnessing}. In other words, AI in finance is no longer only a firm-level productivity issue. It is increasingly a question of market design, institutional control, and financial stability.

Despite this momentum, the academic literature remains conceptually fragmented. Existing studies tend to focus on one of three levels in isolation. The first examines the performance of finance-specific LLMs and reasoning systems. The second studies application domains such as trading, portfolio management, or compliance. The third, often in policy or regulatory work, discusses systemic risk, oversight, and governance. What remains underdeveloped is a unified analytical framework that connects these levels. Yet this connection is precisely where the most important questions arise. The architecture of an AI agent influences what data it sees, how it reasons, how much discretion it is given, and how its outputs are translated into financial action. Those micro-level design choices may in turn shape macro-level outcomes such as information efficiency, liquidity resilience, behavioural correlation, market concentration, and supervisory complexity.

The central argument of the paper is that the systemic impact of AI in finance depends not only on model intelligence but on the architecture, coupling, and governance of agentic systems. In other words, the economic and systemic consequences of AI adoption arise less from the raw capabilities of individual models than from how AI agents are embedded in institutional workflows, technological infrastructures, and market interaction networks.

To analyse this transition, the paper develops an integrated framework for what we call \emph{agentic finance}, defined as financial market settings in which autonomous or semi-autonomous AI systems participate in information processing, decision support, monitoring, or execution in ways that are economically relevant for market outcomes. The analysis is organised around three research questions:

\begin{enumerate}[label=\textbf{RQ\arabic*:}, leftmargin=2.2cm]
    \item What constitutes the architecture of an AI financial agent?
    \item In which financial activities can AI agents meaningfully augment or replace human decision-making?
    \item What systemic implications could arise from the large-scale adoption of AI agents in financial markets?
\end{enumerate}

These questions are motivated by both an empirical reality and a theoretical need. Empirically, financial institutions are already experimenting with AI systems for research support, document analysis, anomaly detection, surveillance, and investment workflows, while more ambitious forms of autonomous trading and agent-based market interaction are moving from prototype to early-stage deployment \citep{yang2023fingpt, xiao2024tradingagents, qian2025whenagentstrade}. Theoretically, however, there is still no sufficiently clear framework for linking the design of AI agents to the structure of financial decision-making and, beyond that, to market-wide consequences.

The paper makes three main contributions. First, it defines agentic finance as a distinct analytical category and proposes a modular four-layer architecture of AI financial agents consisting of data perception, reasoning, strategy generation, and execution with control. Second, it clarifies where AI agents are most plausibly useful in practice by distinguishing bounded augmentation from stronger forms of delegation across trading, portfolio management, risk and compliance, and decentralised finance analytics. Third, it develops the \emph{AFMM} that links micro-level agent design to macro-level market outcomes, including efficiency, liquidity, volatility, herding, concentration risk, and supervisory capacity.

Methodologically, the paper is conceptual but explicitly research-oriented. It is not intended as a purely descriptive survey. Rather, it provides a theory-building framework that can guide future empirical work. The later sections therefore also develop an illustrative empirical application showing how a limited but observable part of the framework can be operationalised using public financial and textual data.

The remainder of the paper is organised as follows. Section~\ref{sec:literature} positions the paper within the existing literature and clarifies its theoretical contribution. Section~\ref{sec:generations} introduces the transition from algorithmic finance to agentic finance and motivates the generational framing of financial AI. Section~\ref{sec:architecture} develops a four-layer architecture of AI financial agents. Section~\ref{sec:afmm} introduces the AFMM, which links agent design parameters to aggregate market outcomes. Section~\ref{sec:applications} maps the principal application domains of AI agents in finance. Section~\ref{sec:governance} discusses governance and regulatory implications, Section~\ref{sec:empirical} outlines an empirical research design for testing the framework, Section~\ref{sec:discussion} discusses broader implications, and Section~\ref{sec:conclusion} concludes.

\section{Literature Positioning and Theoretical Contribution}\label{sec:literature}

This section situates the paper within the rapidly expanding literature on artificial intelligence in finance. Rather than reviewing this literature chronologically, the discussion organises existing work into six thematic streams and clarifies how each contributes to our understanding of AI-driven financial systems. This organisation highlights an important gap: while individual components of agentic finance have been studied extensively, the connections between agent architecture, financial applications, and market-wide consequences remain underdeveloped.

\subsection{Six streams of research on agentic finance}

The relevant literature is broader than the recent wave of financial LLM papers. For the purposes of this paper, it is useful to organise prior work into six streams: (i) agent-based finance and heterogeneous-expectations models, (ii) market microstructure and automated trading, (iii) financial LLMs and reasoning benchmarks, (iv) agent architectures and multi-agent systems, (v) AI in trading and investment management, and (vi) regulation and systemic-risk analysis.

Recent review articles also confirm how quickly this space is maturing. New surveys synthesize the rise of agentic AI architectures in general and AI-agent applications in finance and fintech more specifically \citep{abouali2025agenticai, rizinski2025agentsfinance, aldridge2025agenticfinance}. Those papers are useful reference points for the field. However, their primary objective is synthesis: they review architectures, use cases, and regulatory themes across a rapidly expanding domain. The present paper instead aims to provide a finance-specific conceptual framework that links agent architecture, institutional deployment, governance constraints, and market-level consequences, while also translating those design choices into a stylised market model.

\textbf{Agent-based finance and heterogeneous expectations.}  
A first intellectual antecedent comes from agent-based finance and the broader literature on heterogeneous expectations. Classic surveys and syntheses show how markets populated by heterogeneous agents can generate endogenous volatility, bubbles, crashes, and persistent deviations from benchmark rational-expectations dynamics \citep{lebaron2006agentbased, farmerfoley2009abm, hommes2013behavioral}. This literature is directly relevant to the present paper because the AFMM is ultimately concerned with how diversity, coupling, and interaction among heterogeneous decision systems shape market outcomes. Recent LLM-based market simulations such as ASFM and TwinMarket can be interpreted as a new technological branch of this older heterogeneous-agent tradition, now populated by AI-enabled rather than purely stylised bounded-rational agents \citep{gao2024asfm, yang2025twinmarket}.

\textbf{Market microstructure and automated trading.}  
A second stream comes from market microstructure and the literature on automated and algorithmic trading. This work establishes that automated agents have already altered liquidity provision, price discovery, and intraday volatility long before the arrival of contemporary LLM agents \citep{avellaneda2008hft, hendershott2011liquidity, hasbrouck2018quoting, kirilenko2017flashcrash}. The relevance of this literature is twofold. First, it provides evidence that automation can materially change market quality and market dynamics. Second, it supplies a benchmark against which agentic finance can be compared: the move from algorithmic trading to agentic finance is not the first interaction between automation and markets, but it is a deeper extension from automated execution toward automated workflow coordination.

\textbf{Financial LLMs and reasoning benchmarks.}  
A third stream examines domain adaptation and evaluation of large language models in financial contexts. Early work such as BloombergGPT demonstrated that large-scale domain-specific pretraining can significantly improve performance on financial NLP tasks including sentiment classification, information extraction, and document understanding \citep{wu2023bloomberggpt}. Subsequent work such as FinGPT expanded this approach by proposing open-source pipelines for financial language models and demonstrating their use in financial analysis and forecasting-related workflows \citep{yang2023fingpt}. Benchmark studies such as FinBen further highlight the uneven performance of LLMs across financial reasoning tasks, emphasizing that financial competence cannot be reduced to generic language understanding alone \citep{xie2024finben}. More recent models such as Fin-R1 explore reinforcement-learning-based improvements in structured financial reasoning, particularly for tasks such as compliance analysis and financial question answering \citep{liu2025finr1}.

\textbf{Agent architectures and multi-agent systems.}  
A fourth stream focuses on the architectural foundations of AI agents. General frameworks such as ReAct demonstrate how reasoning and acting can be integrated in language-model workflows \citep{yao2023react}, while generative-agent frameworks illustrate how memory, planning, and environment interaction can be combined in persistent agent-based environments \citep{park2023generative}. Survey work further systematises this design space by analysing memory structures, tool usage, planning mechanisms, and multi-agent coordination \citep{xi2023rise}. Building on these foundations, finance-specific work proposes collaborative multi-agent systems for financial analysis and investment reasoning. FinTeam, for example, models financial analysis as a coordinated interaction between specialised analyst agents responsible for macro interpretation, risk evaluation, and portfolio reasoning \citep{wu2025finteam}.

\textbf{AI in trading and investment management.}  
A fifth stream studies AI deployment in trading, allocation, and investment support. Earlier work in this area includes reinforcement-learning-based trading and portfolio control, as well as early deep-learning approaches to portfolio design and financial prediction \citep{moody2001direct, heaton2017deepportfolios, betancourt2021drlportfolio, lopezdeprado2020ml}. More recent research moves toward LLM-based trading agents and multi-agent trading environments. TradingAgents proposes a collaborative architecture composed of analyst, trader, and risk-management agents \citep{xiao2024tradingagents}, while When Agents Trade introduces a live multi-market benchmark for evaluating LLM trading agents across equities and crypto markets \citep{qian2025whenagentstrade}. Together, these studies show how AI applications are shifting from narrow signal generation toward integrated decision pipelines that include interpretation, strategy construction, and execution support.

\textbf{Regulation and systemic risk.}  
The final stream focuses on governance, regulation, and systemic implications. Reports from the BIS, OECD, and the Financial Stability Board highlight several emerging concerns, including model opacity, concentration of third-party providers, operational resilience, and the potential for correlated machine-mediated behaviour across financial institutions \citep{oecd2023genaifinance, oecd2024regulatory, fsb2024ai, aquilina2025harnessing}. This literature is crucial because it widens the unit of analysis from firm-level productivity to financial stability, supervisory capacity, and market design.

\subsection{Analytical gaps in the existing literature}

Taken together, these six research streams demonstrate that the technological and institutional foundations of AI-driven finance are rapidly evolving. However, three analytical gaps remain particularly important.

\textbf{First, the literature lacks a finance-specific architectural framework for AI agents.}  
Most existing work focuses either on general agent architectures or on specific financial applications. As a result, there is little consensus on how to decompose financial AI agents into modular components that reflect the structure of financial decision-making. Without such an architecture, it is difficult to systematically compare agent designs across institutions or applications.

\textbf{Second, the connection between agent design and financial market outcomes remains underdeveloped.}  
The agent-based finance and market-microstructure literatures already show that heterogeneous automated agents can shape liquidity, volatility, and price discovery. However, these insights are rarely connected to the architecture of contemporary AI systems. Many recent studies still evaluate AI at the level of isolated tasks such as classification, summarisation, or signal generation. Questions related to liquidity provision, volatility amplification, and behavioural correlation therefore remain insufficiently connected to the technical design of modern agentic systems.

\textbf{Third, discussions of systemic risk are often disconnected from the micro-structure of AI systems.}  
Policy-oriented research frequently identifies risks such as model homogeneity, third-party dependencies, and correlated decision-making. Yet these discussions are rarely grounded in a clear model of how agentic systems are constructed, how they process information, and how their outputs translate into financial actions. This limits the ability of both researchers and regulators to link system-level concerns to specific design choices.

\subsection{Theoretical contribution of this paper}

Against this background, the objective of this paper is not simply to review emerging research, but to develop an integrated theoretical framework for what we term \emph{agentic finance}. The central contribution is to bridge the gap between the technical design of AI agents and the economic structure of financial markets.

More specifically, the paper makes three theoretical contributions:

\begin{enumerate}[label=(\roman*), leftmargin=1.2cm]
    \item it proposes a modular four-layer architecture of AI financial agents that decomposes agent systems into data perception, reasoning, strategy generation, and execution with control;

    \item it develops the \emph{AFMM} that links micro-level agent design choices to macro-level market outcomes such as efficiency, liquidity provision, volatility dynamics, herding behaviour, and market concentration;

    \item it formulates a set of mechanism-based propositions that connect technical features of agentic systems—such as model heterogeneity, execution coupling, and infrastructure concentration—to potential financial-system consequences.
\end{enumerate}

The key theoretical move of the paper is therefore to treat AI agents not merely as decision-support tools but as \emph{market actors embedded in shared infrastructures}. Once framed in this way, the relevant analytical variables extend beyond model accuracy or benchmark performance. They include heterogeneity of models and data sources, coupling between decision and execution layers, concentration of technical infrastructure, the degree of delegated autonomy, and the visibility of agent behaviour to supervisors.

By explicitly linking these variables to financial market outcomes, the framework developed in this paper aims to provide a foundation for future empirical research on the economic and systemic implications of agentic financial systems.

This positioning also helps distinguish the paper from adjacent literatures more sharply. Unlike review articles on AI or LLMs in finance, the paper does not primarily synthesise task-level findings. Unlike finance-specific LLM studies, it is not centred on benchmark performance in summarisation, question answering, or prediction. Unlike general agent-architecture papers, it focuses on finance-specific delegation boundaries, market structure, and governance constraints. Unlike the classical agent-based finance literature, it does not begin from stylised behavioural trading rules alone, but from modern AI-enabled decision pipelines embedded in institutions, data infrastructures, and supervisory regimes. The paper's intended contribution is therefore conceptual integration: to connect agent design, institutional deployment, and market-level consequences within a single finance-oriented analytical framework.

The next section places this framework in historical context. Doing so clarifies why the move from algorithmic and machine-learning finance to agentic finance is not simply an incremental improvement in model quality, but a shift in the scope of automation itself.

\section{Three Generations of Financial AI: From Algorithmic Finance to Agentic Finance}\label{sec:generations}

\begin{definition}
\emph{Agentic finance} refers to financial market settings in which autonomous or semi-autonomous AI systems participate in information processing, decision support, monitoring, or execution in ways that are economically relevant for market outcomes.
\end{definition}

To understand why AI agents matter in finance, it is useful to locate them within a broader historical evolution of financial automation. The rise of agentic systems should not be seen as an isolated technological development, but as the latest stage in a longer transition through which finance has progressively automated larger portions of the decision process. In this sense, the contemporary landscape of financial AI can be interpreted as comprising \emph{three generations}: algorithmic finance, machine-learning finance, and agentic finance.

The key distinction across these generations is not simply the sophistication of the underlying model, but the \emph{scope of automation}. Earlier systems automated narrowly defined rules or predictive tasks. More recent systems increasingly automate multi-step workflows involving perception, reasoning, planning, constraint handling, and action. This shift matters because financial decisions are inherently sequential and institutionally constrained: they require not only forecasting, but also interpretation, prioritisation, coordination, execution, and monitoring.

\subsection{First generation: algorithmic finance}

The first generation of financial AI was centred on \emph{algorithmic finance}. Its defining feature was the automation of rule-based execution and market interaction. In this paradigm, systems were designed to follow explicit instructions under pre-specified conditions: for example, splitting orders, optimising execution schedules, or responding to microstructure signals in electronic markets \citep{avellaneda2008hft, hendershott2011liquidity, hasbrouck2018quoting, kirilenko2017flashcrash}. The economic logic of this generation was primarily operational. Automation improved speed, reduced execution costs, and enabled more systematic participation in increasingly electronic markets.

Although algorithmic finance transformed market microstructure, its scope remained limited. These systems did not typically ``understand'' news, reason over changing objectives, or integrate heterogeneous information sources. Their intelligence lay in disciplined execution rather than adaptive interpretation. In other words, the automated object was the \emph{rule}, not the broader decision workflow. This first generation therefore established the infrastructure of automated finance without yet creating genuinely flexible or context-sensitive decision systems.

\subsection{Second generation: machine-learning finance}

The second generation can be described as \emph{machine-learning finance}. Here the main object of automation shifted from execution rules to predictive tasks. Financial institutions increasingly used statistical learning and machine-learning methods for return prediction, volatility forecasting, credit scoring, fraud detection, portfolio analytics, and risk monitoring \citep{moody2001direct, heaton2017deepportfolios, betancourt2021drlportfolio, lopezdeprado2020ml}. Relative to the first generation, this represented a major expansion in analytical capability. Systems were no longer limited to explicit rules; they could infer patterns from data and generate probabilistic assessments that improved classification, ranking, forecasting, and portfolio construction.

Yet machine-learning finance was still largely \emph{model-centric}. Most systems were designed to produce a signal, score, forecast, or recommendation at a specific point in the workflow. Human actors remained responsible for integrating that output into a broader sequence of tasks: gathering contextual information, interpreting significance, checking constraints, and deciding how or whether to act. Even when machine-learning models were highly accurate on narrow tasks, they did not themselves manage the end-to-end process through which financial actions are generated and implemented. Their contribution was therefore substantial, but still partial.

\subsection{Third generation: agentic finance}

The third generation, now emerging, is \emph{agentic finance}. In this generation, AI systems extend beyond isolated prediction or execution tasks and begin to operate across a larger portion of the financial workflow. Advances in large language models, retrieval systems, planning modules, memory architectures, and tool-using agents have made it increasingly feasible to build systems that can absorb heterogeneous inputs, reason across intermediate steps, coordinate external tools, maintain state over time, and recommend or initiate actions in sequential environments \citep{yao2023react, park2023generative, xi2023rise}.

What distinguishes agentic finance from earlier generations is therefore not only improved intelligence, but \emph{workflow integration}. An agent is not merely a forecasting model or an execution algorithm. It is a goal-directed system that can transform inputs into market-relevant decision objects under explicit objectives and constraints. In financial contexts, those decision objects may include trade ideas, portfolio reallocations, anomaly alerts, hedging proposals, compliance flags, or monitoring outputs. The critical point is that the system participates in the chain that links information to action.

This broader scope also changes the risk profile of AI in finance. In algorithmic finance, the main concerns centred on execution speed, market impact, and microstructure instability. In machine-learning finance, concerns expanded to include model risk, overfitting, and interpretability. In agentic finance, these issues remain relevant, but they are joined by additional concerns related to autonomy, coordination, infrastructure dependence, control loss, and the possibility of correlated system behaviour across institutions. As a result, the analytical challenge is no longer simply to ask whether AI improves local task performance. It is to ask how different forms of agentic design may reshape financial workflows and, ultimately, market dynamics.

\subsection{From model-centric automation to workflow-centric automation}

The transition across the three generations can be summarised as a shift from \emph{rule automation}, to \emph{prediction automation}, to \emph{workflow automation}. This framing is important because it clarifies why agentic finance deserves separate analytical treatment. If AI systems increasingly operate across multiple stages of financial decision-making, then the relevant unit of analysis is no longer the isolated model, but the \emph{agent system} as a structured combination of data perception, reasoning, strategy formation, and execution.

\begin{figure}[ht]
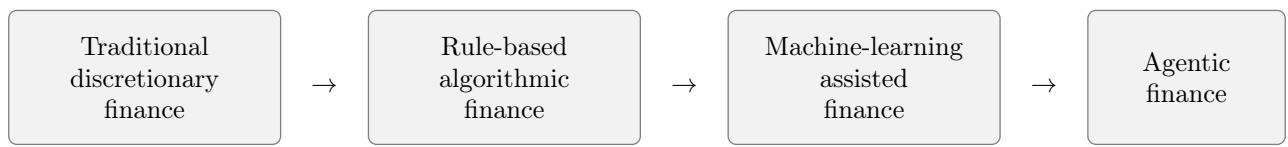

\centering
\small

\makebox[\textwidth][c]{

\begin{tcolorbox}[width=0.22\textwidth,height=1.8cm,
colback=black!5,colframe=black!50,valign=center]
\centering Traditional\\discretionary\\finance
\end{tcolorbox}

\hspace{0.01\textwidth}
\raisebox{0.7cm}{$\rightarrow$}
\hspace{0.01\textwidth}

\begin{tcolorbox}[width=0.22\textwidth,height=1.8cm,
colback=black!5,colframe=black!50,valign=center]
\centering Rule-based\\algorithmic\\finance
\end{tcolorbox}

\hspace{0.01\textwidth}
\raisebox{0.7cm}{$\rightarrow$}
\hspace{0.01\textwidth}

\begin{tcolorbox}[width=0.22\textwidth,height=1.8cm,
colback=black!5,colframe=black!50,valign=center]
\centering Machine-learning\\assisted\\finance
\end{tcolorbox}

\hspace{0.01\textwidth}
\raisebox{0.7cm}{$\rightarrow$}
\hspace{0.01\textwidth}

\begin{tcolorbox}[width=0.16\textwidth,height=1.8cm,
colback=black!5,colframe=black!50,valign=center]
\centering Agentic\\finance
\end{tcolorbox}

}

\caption{Three generations of financial AI: from rule-based execution to agentic financial systems.}
\label{fig:evolution}

\end{figure}

Figure~\ref{fig:evolution} illustrates this progression. The figure is intentionally stylised: in practice, the three generations overlap, and institutions often combine them within the same workflow. A modern financial system may still rely on algorithmic execution, machine-learning forecasts, and agentic reasoning simultaneously. Nevertheless, the generational framing is analytically useful because it highlights a real transformation in the locus of automation. The central question is no longer whether AI can assist a narrow financial task, but how increasingly agentic systems alter the structure of financial decision-making itself.

This conceptual distinction motivates the next section. If agentic finance represents a shift toward workflow-centric automation, then the first task is to specify the internal structure of such systems. The following section therefore develops a modular architecture of AI financial agents and identifies the components through which they perceive information, reason under constraints, generate financial decision objects, and interact with execution and control infrastructures.

\section{A Four-Layer Architecture of AI Financial Agents}\label{sec:architecture}

The emergence of agentic finance requires a more explicit description of how AI systems are structured inside financial decision workflows. Existing research often focuses on isolated components such as forecasting models, language models, or execution algorithms. However, once AI systems operate across multiple stages of financial decision-making, the relevant unit of analysis is no longer the individual model but the \emph{agent system} that integrates perception, reasoning, strategy formation, and execution.

This section therefore introduces a modular architecture for AI financial agents. The framework decomposes agent systems into four layers: \emph{data perception}, \emph{reasoning}, \emph{strategy generation}, and \emph{execution with control}. Each layer corresponds to a distinct stage in the transformation of raw information into market-relevant actions.

The purpose of this architecture is not to prescribe a single implementation. Rather, it provides a conceptual decomposition that allows different institutional designs to be analysed within a common framework. The key analytical advantage of this approach is that it separates the \emph{cognitive functions} of financial AI from the \emph{institutional controls} that govern their use.

\begin{figure}[t]
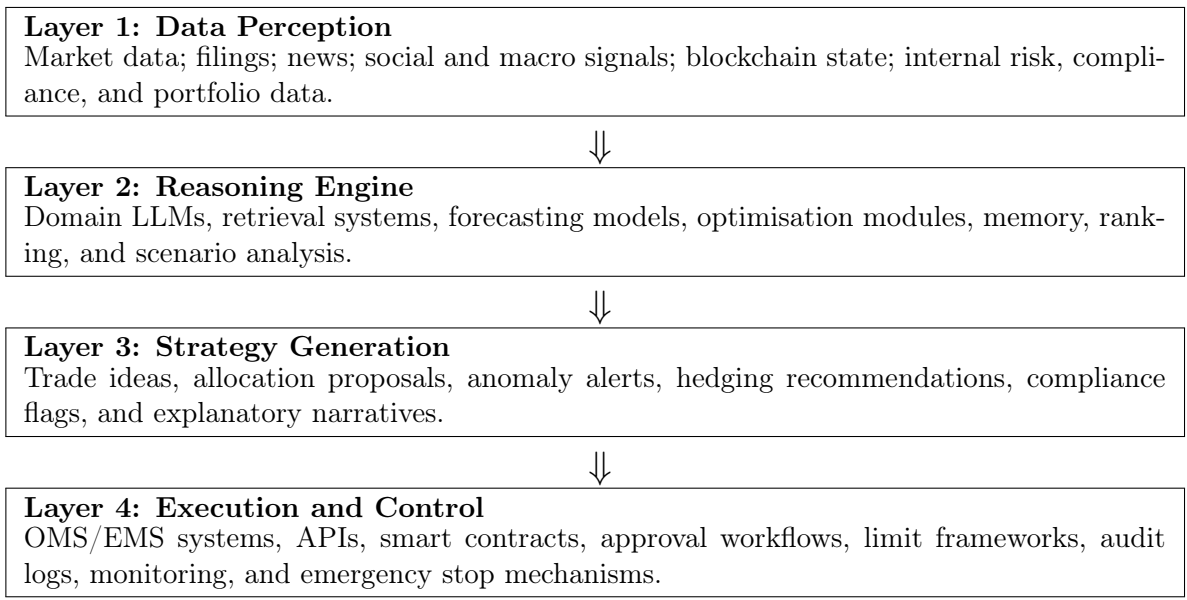

\centering
\layerbox{Layer 1: Data Perception}{Market data; filings; news; social and macro signals; blockchain state; internal risk, compliance, and portfolio data.}

\vspace{0.4em}
\Large$\Downarrow$
\normalsize

\layerbox{Layer 2: Reasoning Engine}{Domain LLMs, retrieval systems, forecasting models, optimisation modules, memory, ranking, and scenario analysis.}

\vspace{0.4em}
\Large$\Downarrow$
\normalsize

\layerbox{Layer 3: Strategy Generation}{Trade ideas, allocation proposals, anomaly alerts, hedging recommendations, compliance flags, and explanatory narratives.}

\vspace{0.4em}
\Large$\Downarrow$
\normalsize

\layerbox{Layer 4: Execution and Control}{OMS/EMS systems, APIs, smart contracts, approval workflows, limit frameworks, audit logs, monitoring, and emergency stop mechanisms.}

\caption{A four-layer architecture for AI agents in financial markets.}
\label{fig:architecture}
\end{figure}

\subsection{Layer 1: Data perception}

The first layer concerns how financial agents perceive and organise information. Financial environments are characterised by unusually heterogeneous data structures. Relevant signals may include high-frequency market prices, order-book dynamics, company disclosures, regulatory filings, earnings-call transcripts, macroeconomic releases, social-media signals, blockchain transaction flows, and internal portfolio data.

The primary challenge at this stage is not only data ingestion but also \emph{data governance}. Financial information differs in latency, reliability, legal access rights, and economic interpretation. A market data feed may update thousands of times per second, whereas accounting information may change quarterly. Similarly, some information sources are public, while others are proprietary or subject to regulatory restrictions.

Consequently, the perception layer must perform several tasks simultaneously: data normalisation, timestamp alignment, provenance tracking, and access control. These functions determine which signals an agent can observe and how quickly it can respond to them. In practice, the design of this layer strongly shapes the informational advantage of financial agents, because differences in data pipelines often translate directly into differences in market responsiveness.

\subsection{Layer 2: Reasoning engine}

The second layer transforms perceived information into intermediate beliefs, forecasts, and explanations. This layer constitutes the cognitive core of the agent system.

In modern financial AI systems, the reasoning layer is rarely implemented as a single model. Instead, it typically combines multiple analytical components. Large language models provide capabilities for interpreting unstructured text such as filings, news, and regulatory documents. Retrieval systems enable agents to access historical context or domain knowledge. Statistical models and machine-learning algorithms generate forecasts of returns, volatility, credit risk, or liquidity conditions. Optimisation modules translate these forecasts into feasible portfolio or trading configurations.

Recent developments in language-model agents further extend these capabilities by enabling sequential reasoning, planning, and tool usage \citep{yao2023react, park2023generative, xi2023rise}. In such systems, the reasoning engine can break complex tasks into intermediate steps, query external data sources, evaluate alternative hypotheses, and maintain internal memory across interactions. For financial decision-making, this hybrid design is particularly important because no single model type is sufficient to address the full range of analytical tasks involved.

\subsection{Layer 3: Strategy generation}

The third layer converts analytical outputs into \emph{decision objects}. A decision object is a structured representation of a potential financial action together with the constraints and reasoning that justify it. Examples include trade proposals, portfolio reallocations, hedging strategies, anomaly alerts, compliance flags, or monitoring signals.

This concept is important because it clarifies the role of AI agents in financial workflows. Rather than producing isolated predictions, agent systems generate decision objects that integrate information, reasoning, and institutional constraints. For example, a trading agent may produce a trade idea together with a confidence score, risk estimate, liquidity assessment, and explanation derived from recent news or macro events.

Strategy generation is therefore the stage at which analytical insight becomes economically meaningful. At this stage, the agent must reconcile forecasts with practical constraints such as portfolio mandates, liquidity limits, regulatory requirements, and transaction costs. The design of this layer determines whether the agent functions primarily as a research assistant, a portfolio co-pilot, or a more autonomous decision system.

\subsection{Layer 4: Execution and control}

The final layer connects decision objects to financial infrastructure. This layer includes order-management systems, execution-management systems, exchange APIs, smart-contract interactions, and other operational interfaces through which financial actions are implemented.

However, execution is not purely technical. It is also the point at which \emph{institutional control} becomes critical. Financial institutions must ensure that AI-driven decisions remain consistent with risk limits, legal obligations, and governance frameworks. As a result, the execution layer typically incorporates approval workflows, position limits, audit trails, monitoring systems, and emergency stop mechanisms.

The presence of these controls means that most real-world financial agents operate under \emph{bounded autonomy}. Agents may generate recommendations or initiate actions within predefined limits, but human oversight remains responsible for escalation decisions and exceptional situations. From a systemic perspective, the design of the execution layer is particularly important because it determines how quickly agent decisions propagate into market activity.

\subsection{Design dimensions}

While the four-layer architecture describes the internal structure of financial agents, the broader theoretical implications of agentic finance depend on several design dimensions that cut across these layers. These dimensions serve as the key primitives in the \emph{Agentic Financial Market Model (AFMM)} developed later in the paper.

\begin{itemize}[leftmargin=1.2cm]

\item \textbf{Autonomy depth.}  
The extent to which agents can act without human approval. Some systems only generate analytical summaries, while others can initiate trades or rebalance portfolios within predefined constraints.

\item \textbf{Model heterogeneity.}  
The diversity of models, prompts, and data sources used by different agents. Higher heterogeneity may increase informational diversity, while strong homogeneity may create correlated behaviour across institutions.

\item \textbf{Execution coupling.}  
The degree to which multiple agents respond to similar signals on similar timescales. Strong coupling may increase the speed of price discovery but can also amplify market volatility.

\item \textbf{Infrastructure concentration.}  
The extent to which agents rely on shared cloud platforms, model providers, APIs, or data vendors. High concentration introduces potential systemic dependencies.

\item \textbf{Supervisory observability.}  
The ability of internal control teams or regulators to reconstruct the decision path of AI agents, including inputs, intermediate reasoning steps, and executed actions.

\end{itemize}

Together, these dimensions connect micro-level system design to macro-level financial outcomes. They provide the analytical bridge between the architecture described in this section and the market-level framework introduced in the next section.

\section{The Agentic Financial Market Model (AFMM)}\label{sec:afmm}

The preceding sections introduced two key elements of agentic finance: the generational transition toward workflow-level automation and the internal architecture of AI financial agents. The next step is to connect these micro-level design features to market-level outcomes. This section therefore introduces the \emph{Agentic Financial Market Model (AFMM)}, a conceptual framework that links the distribution and interaction of AI agents to financial market dynamics.

The AFMM can be interpreted as a stylised agent-based representation of financial markets populated by heterogeneous AI-enabled decision systems. It is not a calibrated simulation model and it is not a fully solved equilibrium model. Rather, it is an analytical skeleton designed to isolate the mechanisms through which heterogeneous agent architectures may shape market outcomes.

The objective of the AFMM is not to produce a fully specified equilibrium model, but rather to identify the key mechanisms through which agentic systems influence market behaviour. In particular, the model focuses on how differences in autonomy, model diversity, execution synchronisation, infrastructure dependencies, and supervisory control interact to shape market efficiency, liquidity resilience, volatility, and systemic risk.

\subsection{Model intuition}

The central intuition of the AFMM is that the market impact of AI agents depends less on their individual intelligence and more on how they are \emph{distributed, coupled, and governed} across the financial system.

A market populated by heterogeneous agents drawing on diverse models, data sources, and objectives may enhance price discovery by increasing informational diversity. In contrast, a market in which many institutions rely on highly similar agents, trained on overlapping datasets and connected to the same technological infrastructure, may produce correlated responses to common signals. Under such conditions, even locally optimal decisions can aggregate into market-wide instability.

The AFMM therefore shifts the focus from individual model performance to the \emph{population structure of agents}. In this perspective, financial markets increasingly resemble ecosystems of interacting human and machine decision-makers embedded in shared infrastructures.

\begin{figure}[ht]
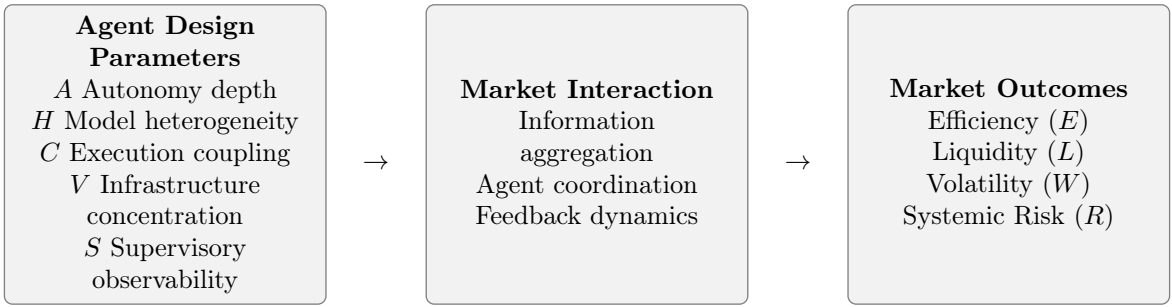

\centering
\small

\makebox[\textwidth][c]{

\begin{tcolorbox}[
width=0.26\textwidth,
height=4cm,
colback=black!5,
colframe=black!50,
valign=center,
top=3mm,
bottom=3mm]
\centering
\textbf{Agent Design Parameters}\\
$A$ Autonomy depth\\
$H$ Model heterogeneity\\
$C$ Execution coupling\\
$V$ Infrastructure concentration\\
$S$ Supervisory observability
\end{tcolorbox}

\hspace{0.015\textwidth}
\raisebox{1.8cm}{$\rightarrow$}
\hspace{0.015\textwidth}

\begin{tcolorbox}[
width=0.26\textwidth,
height=4cm,
colback=black!5,
colframe=black!50,
valign=center,
top=3mm,
bottom=3mm]
\centering
\textbf{Market Interaction}\\
Information aggregation\\
Agent coordination\\
Feedback dynamics
\end{tcolorbox}

\hspace{0.015\textwidth}
\raisebox{1.8cm}{$\rightarrow$}
\hspace{0.015\textwidth}

\begin{tcolorbox}[
width=0.26\textwidth,
height=4cm,
colback=black!5,
colframe=black!50,
valign=center,
top=3mm,
bottom=3mm]
\centering
\textbf{Market Outcomes}\\
Efficiency ($E$)\\
Liquidity ($L$)\\
Volatility ($W$)\\
Systemic Risk ($R$)
\end{tcolorbox}

}

\caption{Conceptual structure of the Agentic Financial Market Model (AFMM).}
\label{fig:afmm}

\end{figure}

Figure~\ref{fig:afmm} summarises the conceptual structure of the AFMM. 
Agent design parameters determine how agents process information and 
interact in markets, which in turn shapes aggregate market outcomes.

\subsection{Agent population and primitives}

Consider a market with $N$ participants indexed by $i = 1,\dots,N$. A subset of these participants employ AI agents as part of their financial decision workflow. For each AI-enabled participant, define the following design variables:

\begin{align}
A_i &\in [0,1] && \text{autonomy depth}, \\
H_i &\in [0,1] && \text{model and data heterogeneity}, \\
C_i &\in [0,1] && \text{execution coupling to common signals}, \\
V_i &\in [0,1] && \text{vendor or infrastructure concentration exposure}, \\
S_i &\in [0,1] && \text{supervisory observability and control capacity}.
\end{align}

These variables correspond directly to the design dimensions introduced in the architecture framework of Section~\ref{sec:architecture}.

\begin{itemize}[leftmargin=1.2cm]

\item \textbf{Autonomy depth ($A_i$).}  
The degree to which an agent can initiate or execute actions without human intervention.

\item \textbf{Model heterogeneity ($H_i$).}  
The diversity of models, data pipelines, prompts, and objectives across agents.

\item \textbf{Execution coupling ($C_i$).}  
The extent to which agents respond to similar signals on similar timescales.

\item \textbf{Infrastructure concentration ($V_i$).}  
The dependence of agents on common upstream providers such as cloud platforms, model vendors, data feeds, or middleware.

\item \textbf{Supervisory observability ($S_i$).}  
The ability of institutions or regulators to reconstruct and intervene in the decision paths of agents.

\end{itemize}

Let market-level outcomes be represented by the following variables:

\begin{align}
E &= \text{informational efficiency}, \\
L &= \text{liquidity resilience}, \\
W &= \text{volatility or instability}, \\
R &= \text{systemic risk}.
\end{align}

These outcomes capture the primary dimensions along which agentic finance may influence market structure.

\subsection{Agent decision function}

To connect agent design parameters to market behaviour, we introduce a stylised representation of the decision process of AI-enabled financial agents.

At each time $t$, agent $i$ observes a vector of market signals
\[
X_t = (P_t, Z_t, N_t),
\]
where $P_t$ denotes asset prices, $Z_t$ represents macro or market state variables, and $N_t$ captures textual or informational signals such as news or disclosures.

Given these inputs, the agent produces a decision object $D_{i,t}$ through a decision function
\begin{equation}
D_{i,t} = g_i(X_t; A_i, H_i, C_i, V_i, S_i),
\end{equation}
where $g_i(\cdot)$ represents the internal reasoning pipeline of the agent. This function abstracts the layered architecture described earlier in the paper, combining data perception, reasoning, strategy generation, and execution constraints.

The parameters $(A_i,H_i,C_i,V_i,S_i)$ influence the properties of this decision rule. Higher autonomy depth ($A_i$) allows the agent to initiate actions with fewer human approval constraints. Greater heterogeneity ($H_i$) reflects diversity in models, training data, and objectives across agents. Execution coupling ($C_i$) captures the extent to which agents respond to common signals or triggers. Infrastructure concentration ($V_i$) represents reliance on shared technological providers, while supervisory observability ($S_i$) captures institutional control capacity and monitoring visibility.

The output $D_{i,t}$ may correspond to various financial actions, including trade instructions, portfolio rebalancing proposals, risk alerts, or monitoring signals.

Because financial agents do not always execute their recommendations directly, it is useful to distinguish between a \emph{decision object} and its realised market action. Let the realised action be
\begin{equation}
q_{i,t} = \Lambda_i(D_{i,t}; A_i, S_i),
\end{equation}
where $\Lambda_i(\cdot)$ maps decision objects into executed actions through approval rules, limits, supervisory checks, and other institutional controls. This distinction is important because two agents may generate similar internal recommendations while differing sharply in the speed or scale with which those recommendations reach the market.

To capture the possibility of correlated machine-mediated behaviour, define the average similarity of realised actions as
\begin{equation}
\rho_t = \frac{2}{N(N-1)} \sum_{i<j} \mathrm{corr}(q_{i,t}, q_{j,t}),
\end{equation}
where higher $\rho_t$ corresponds to greater synchronisation in realised actions. In the spirit of the AFMM, $\rho_t$ should be increasing in execution coupling and common infrastructure dependence, and decreasing in model heterogeneity. In reduced form, this implies
\begin{equation}
\rho_t = \psi(\bar{C}_t, 1-\bar{H}_t, \bar{V}_t).
\end{equation}

\subsection{Market aggregation}

Market outcomes emerge from the aggregation of realised rather than merely recommended actions. The aggregate market action can therefore be written as

\begin{equation}
Q_t = \sum_{i=1}^{N} w_i q_{i,t},
\end{equation}

where $w_i$ represents the market weight or trading capacity of participant $i$.

Asset prices and liquidity conditions evolve as a function of the aggregate order flow:

\begin{equation}
P_{t+1} = F(P_t, Q_t, X_t),
\end{equation}

where $F(\cdot)$ captures the market impact of aggregated decisions under prevailing market conditions.

More generally, the market state evolves according to
\begin{equation}
X_{t+1} = T(X_t, Q_t, \varepsilon_{t+1}),
\end{equation}
where $\varepsilon_{t+1}$ denotes exogenous news and liquidity shocks. The AFMM is therefore explicitly dynamic: agent design affects realised actions, realised actions affect market states, and updated market states feed back into future agent decisions.

This representation highlights a key feature of agentic finance: market dynamics depend not only on individual decision quality but also on the \emph{correlation structure of agent decisions}. When many agents rely on similar models, signals, or infrastructure, the resulting decisions may become highly correlated, amplifying market reactions.

In this sense, the AFMM focuses on how the joint distribution of agent design parameters $(A_i,H_i,C_i,V_i,S_i)$ shapes the statistical properties of the aggregate decision process $Q_t$ through both the scale of realised actions and the correlation structure summarised by $\rho_t$.

\subsection{Market interaction mechanisms}

The AFMM assumes that market outcomes emerge from the interaction between the cross-sectional distribution of agent design parameters and the prevailing market environment.

In reduced form, market-level outcomes can be written as:

\begin{align}
E &= f_E(\bar{A}, \bar{H}, \bar{S}, X), \\
L &= f_L(\bar{A}, \bar{H}, \rho, \bar{V}, X), \\
W &= f_W(\bar{A}, \rho, \bar{V}, 1-\bar{S}, X), \\
R &= f_R(\bar{A}, \rho, \bar{V}, 1-\bar{H}, 1-\bar{S}, X),
\end{align}

where bars denote market-level averages or concentration-weighted aggregates and $X$ represents exogenous market conditions such as macroeconomic shocks or liquidity stress.

The qualitative logic underlying these relationships can be summarised as follows.

First, greater autonomy increases the speed and scale of market reactions. While this may enhance efficiency under normal conditions, it may also amplify rapid feedback loops during periods of stress.

Second, greater heterogeneity across models and data pipelines increases informational diversity. This tends to improve price discovery by ensuring that agents interpret signals differently rather than converging on a single dominant narrative.

Third, strong execution coupling increases the probability of synchronous reactions. When many agents respond to similar signals at similar times, market reactions may become highly correlated, pushing $\rho_t$ upward.

Fourth, dependence on common technological infrastructure introduces common-mode failure risk. If many institutions rely on the same upstream providers, outages or model failures may propagate simultaneously across the system and increase the effective similarity of realised actions even when institutions appear organizationally distinct.

Finally, strong supervisory observability acts as a stabilising force by enabling effective monitoring, intervention, and governance.

\subsection{Core propositions}

The AFMM generates several qualitative propositions about the behaviour of agentic financial markets.

\begin{proposition}
Holding other factors constant, greater heterogeneity in models, data sources, and decision objectives increases the likelihood that AI agents contribute positively to price discovery rather than merely reinforce common narratives.
\end{proposition}

\begin{proposition}
Holding other factors constant, stronger execution coupling increases the probability that local decision improvements translate into market-wide herding, episodic liquidity withdrawal, and volatility amplification.
\end{proposition}

\begin{proposition}
Systemic risk rises non-linearly when increases in autonomy depth and infrastructure concentration outpace improvements in supervisory observability and institutional control capacity.
\end{proposition}

\subsection{From theoretical propositions to empirical hypotheses}

Although the AFMM is primarily conceptual, its structure naturally suggests empirical hypotheses that can be tested using market data or simulation environments.

\begin{hypothesis}
Public capability disclosures about agentic AI generate heterogeneous market repricing across firms whose revenues or operating costs are differentially exposed to workflow substitution and legacy modernization.
\end{hypothesis}

\begin{hypothesis}
Agentic systems generate larger performance gains in information-intensive analytical tasks than in fully autonomous real-time trading tasks where execution constraints dominate.
\end{hypothesis}

\begin{hypothesis}
Common upstream technological dependencies are associated with higher cross-institution similarity in AI-generated market responses.
\end{hypothesis}

\subsection{Why the AFMM matters}

The AFMM strengthens the theoretical contribution of this paper in two ways. First, it transforms the architectural framework of Section~\ref{sec:architecture} into a mechanism-based explanation of market dynamics. Rather than treating AI adoption as a purely technological phenomenon, the model highlights the systemic consequences of agent design choices.

Second, the AFMM creates a bridge between conceptual analysis and empirical research. By expressing market outcomes as functions of agent design parameters, the framework provides a structure within which future empirical studies can test how different configurations of agentic finance influence market efficiency, volatility, and systemic stability.

\section{Applications of AI Agents in Finance}\label{sec:applications}

While the AFMM provides a conceptual framework for understanding how AI agents may influence market outcomes, the practical significance of agentic finance ultimately depends on how these systems are deployed in real financial activities. In practice, AI agents rarely operate as fully autonomous decision-makers. Instead, they function as \emph{bounded agents} embedded within institutional workflows, where human oversight, regulatory constraints, and risk controls remain central.

This section maps the architecture and theoretical mechanisms developed earlier to several major application domains. The goal is not to predict a single dominant trajectory of adoption, but to illustrate how agentic systems may augment financial decision processes across trading, portfolio management, risk monitoring, and decentralised finance ecosystems.

\subsection{Autonomous trading}

Trading is the most natural domain for the deployment of AI agents because it already relies heavily on automated infrastructure such as algorithmic execution systems, market data pipelines, and order management platforms. However, the transition from algorithmic trading to agentic trading primarily concerns \emph{workflow integration} rather than complete autonomy.

In traditional algorithmic trading systems, strategies are typically predefined and executed through rule-based engines. AI agents extend this architecture by integrating multiple tasks within a single decision pipeline. For example, an agent may simultaneously monitor news flows, interpret macroeconomic signals, update forecasts, and generate candidate trade instructions. The agent can then pass proposed orders through execution algorithms subject to risk limits and compliance filters.

In practice, most institutions are likely to adopt \emph{bounded autonomy} models in which AI agents assist with signal synthesis, order preparation, and post-trade analysis rather than initiating large-scale autonomous trading. Such designs reduce operational risk while still allowing agents to improve reaction speed and information processing capacity.

Recent developments in LLM-based trading frameworks illustrate this shift toward agentic workflows. Multi-agent trading environments allow different agents to specialise in forecasting, portfolio construction, and risk monitoring, while interacting through structured communication protocols \citep{yang2023fingpt, xiao2024tradingagents}. These architectures resemble distributed research teams rather than single autonomous traders.

Within the AFMM framework, trading applications are particularly sensitive to the parameters of autonomy depth ($A$) and execution coupling ($C$). High autonomy combined with strong signal coupling could amplify synchronous reactions across institutions, potentially contributing to episodic volatility events similar to those observed in earlier algorithmic trading episodes \citep{kirilenko2017flashcrash}.

\subsection{Portfolio management}

In portfolio management, the most plausible role for AI agents is not full automation but \emph{research augmentation}. Asset allocation decisions involve complex trade-offs between macroeconomic expectations, valuation models, institutional mandates, and client preferences. These tasks require structured reasoning, scenario construction, and narrative explanation—areas where large language models and agentic workflows can provide meaningful assistance.

AI agents can support portfolio managers by synthesising large volumes of heterogeneous information, including macroeconomic releases, corporate disclosures, earnings calls, and geopolitical developments. Instead of producing simple predictions, agents can generate structured research outputs such as investment memos, scenario analyses, and portfolio stress tests.

Another promising application lies in scenario generation and probabilistic reasoning. Agents can construct multiple hypothetical macroeconomic trajectories and evaluate how different asset classes may perform under alternative policy or market conditions. Such capabilities align closely with the reasoning and planning modules typical of agentic architectures.

Importantly, the fiduciary nature of asset management imposes strong constraints on autonomy depth. Most portfolio decisions require explicit human approval, meaning that agents primarily function as \emph{co-pilot systems} rather than independent allocators. In the AFMM framework, this corresponds to relatively low values of $A$ but potentially high values of informational heterogeneity ($H$), which may enhance the diversity of analytical perspectives available to portfolio managers.

\subsection{Risk monitoring and compliance}

Risk management and compliance monitoring represent one of the most immediately impactful domains for AI agents in finance. Financial institutions already operate extensive monitoring systems for market risk, credit exposure, operational risk, and regulatory compliance. However, these systems often rely on static rule sets or manually maintained indicators.

Agentic systems can enhance these monitoring frameworks by continuously analysing transaction flows, market data, and internal control signals in real time. AI agents can detect anomalous patterns, flag suspicious transactions, or identify emerging liquidity stress before traditional indicators trigger alerts.

From a regulatory perspective, the ability of AI agents to process large volumes of heterogeneous information may improve the timeliness and granularity of risk detection \citep{aldasoro2024intelligent}. For example, agents could monitor market microstructure indicators, news flows, and funding conditions simultaneously to identify early signs of systemic stress.

However, the deployment of AI agents in compliance settings raises important governance challenges. False positives may generate excessive alerts, while opaque decision processes may complicate accountability. These concerns highlight the importance of supervisory observability ($S$) within the AFMM. Institutions must ensure that AI-generated alerts can be audited, interpreted, and challenged by human supervisors.

More broadly, AI agents may play an important role in the supervision of financial markets themselves. Regulators are increasingly exploring AI tools to monitor trading behaviour, detect market manipulation, and analyse systemic risk indicators across multiple institutions \citep{crisanto2024regulating, aquilina2025harnessing}.

\subsection{DeFi and on-chain intelligence}

Decentralised finance (DeFi) ecosystems provide a particularly fertile environment for AI agents because financial activity on blockchains generates large volumes of publicly observable data. Transactions, liquidity flows, governance proposals, and smart contract interactions are recorded on-chain, creating rich datasets for automated analysis.

AI agents can analyse blockchain data to trace wallet activity, detect unusual trading behaviour, or monitor governance dynamics in decentralised protocols. For example, agents may track large liquidity movements across decentralised exchanges, identify emerging arbitrage opportunities, or evaluate the security risks of newly deployed smart contracts.

Another important application lies in the monitoring of decentralised governance processes. AI agents can analyse proposal discussions, voting behaviour, and token distributions to assess the likelihood of governance capture or coordinated manipulation.

The integration of AI agents with DeFi protocols also enables new forms of automated financial services. Agents may dynamically allocate liquidity across decentralised lending platforms, manage collateral positions, or monitor liquidation risks in real time. Because DeFi protocols operate continuously and transparently, such agents can interact directly with smart contracts without the operational frictions typical of traditional financial infrastructure.

Within the AFMM framework, DeFi applications illustrate the interaction between high autonomy depth ($A$) and fragmented infrastructure. While blockchain transparency increases observability, fragmented liquidity across multiple protocols may create new forms of execution coupling and systemic fragility.

Table~\ref{tab:applications} summarises the main application domains of AI agents in finance, highlighting typical tasks, autonomy levels, and institutional constraints. Together, these examples illustrate that the most significant impact of agentic finance may arise not from fully autonomous decision systems but from the gradual integration of AI agents into existing financial workflows.

\begin{table}[ht]
\centering
\small
\renewcommand{\arraystretch}{1.2}

\begin{tabularx}{\linewidth}{
>{\raggedright\arraybackslash}p{0.20\linewidth}
>{\raggedright\arraybackslash}p{0.24\linewidth}
>{\raggedright\arraybackslash}p{0.21\linewidth}
Y}

\toprule
\textbf{Use case} & \textbf{Typical agent tasks} & \textbf{Likely autonomy level} & \textbf{Main constraints} \\
\midrule

Autonomous trading & Signal synthesis, news interpretation, order preparation, post-trade analysis & Medium in narrow mandates & Latency, slippage, best execution, manipulation risk \\

Portfolio management & Macro synthesis, memo generation, allocation suggestions, scenario analysis & Low to medium & Fiduciary duty, mandate constraints, explainability \\

Risk and compliance & Alerting, anomaly detection, policy mapping, transaction surveillance & Medium for triage & False positives, accountability, due process \\

DeFi intelligence & Wallet tracing, governance monitoring, liquidity mapping, contract-event screening & Medium & Data noise, oracle risk, fragmented venues \\

\bottomrule
\end{tabularx}

\caption{Application map for AI agents in finance.}
\label{tab:applications}

\end{table}

Across these domains, a common pattern emerges. The immediate economic impact of agentic finance is likely to come less from fully autonomous market actors than from the gradual reconfiguration of research, monitoring, execution, and control workflows inside existing institutions. That pattern is precisely why governance becomes central. Once agentic systems are embedded in material financial processes, the key question is no longer whether they improve local tasks, but whether their delegation boundaries, infrastructure dependencies, and oversight mechanisms are robust enough for system-wide adoption.

\section{Governance and Regulation of Agentic Finance}\label{sec:governance}

The emergence of agentic finance raises important governance and regulatory questions. While previous waves of financial automation primarily affected trading infrastructure, AI agents operate at the level of decision workflows. As a result, they may influence not only market efficiency but also the transparency, accountability, and stability of financial systems.

The governance problem in agentic finance is therefore not whether AI should be used, but under what institutional conditions delegation remains safe, auditable, and socially acceptable. This section discusses the governance challenges associated with agentic systems, outlines potential policy responses, and derives a set of governance principles implied by the AFMM framework.

\subsection{Governance challenges of agentic finance}

The increasing deployment of AI agents introduces several governance challenges for financial institutions and regulators.

First, the use of autonomous or semi-autonomous agents complicates the attribution of responsibility. When decisions emerge from interactions among multiple models, data sources, and optimisation modules, identifying the origin of an error or unexpected market outcome becomes more difficult. This raises important questions regarding accountability, auditability, and model governance.

Second, AI agents may increase the opacity of financial decision processes. Complex model pipelines and adaptive learning systems can make it difficult for supervisors to reconstruct the reasoning behind specific decisions. Regulatory frameworks therefore increasingly emphasise explainability and traceability in AI-enabled financial systems \citep{fsb2024ai}.

Third, the concentration of technological infrastructure introduces new forms of systemic risk. Many financial institutions rely on common upstream providers for cloud computing, foundation models, data services, or trading infrastructure. If a large number of AI agents depend on similar models or APIs, failures in these shared components may propagate across the financial system.

Fourth, AI agents may accelerate the speed of market reactions. While faster information processing can improve efficiency, it may also amplify feedback loops during periods of stress. These dynamics resemble earlier concerns about high-frequency trading, but may operate at a broader scale because agentic systems integrate information processing, reasoning, and decision-making in a single pipeline.

Within the AFMM framework, these governance issues correspond directly to the model parameters of infrastructure concentration ($V$), supervisory observability ($S$), and execution coupling ($C$). Effective governance therefore requires institutional mechanisms that maintain sufficient observability and control as agent autonomy increases.

\subsection{Policy responses and supervisory strategies}

Regulators and international organisations have begun to address the implications of AI adoption in financial markets. Recent policy discussions emphasise several supervisory strategies.

In Europe, several recent regulations provide partial legal anchors for this agenda even though none was designed specifically for agentic finance. The Digital Operational Resilience Act (DORA) strengthens requirements around operational resilience, ICT risk management, and oversight of critical third-party technology providers \citep{eu2022dora}. The Markets in Crypto-Assets Regulation (MiCA) shows how digital-asset infrastructures are increasingly being brought within explicit governance, disclosure, and conduct frameworks \citep{eu2023mica}. The EU AI Act extends this trajectory by imposing risk-based obligations around high-impact AI systems, documentation, monitoring, and human oversight \citep{eu2024aiact}. Recent practitioner-oriented work on agentic investment firms points in a similar direction by stressing that architectural delegation, compliance automation, and supervisory integration must be designed together rather than treated as separate afterthoughts \citep{joshi2026agenticrias}. Taken together, these frameworks do not yet solve the governance problem of agentic finance, but they illustrate the direction of travel: resilience, traceability, outsourcing dependence, and human accountability are moving from soft principles toward harder institutional requirements.

First, institutions are expected to implement robust model governance frameworks that document training data, model architecture, and decision logic. Such documentation helps ensure that AI-driven decisions remain auditable and accountable.

Second, supervisory authorities may require financial firms to maintain \emph{human-in-the-loop} controls for high-impact financial decisions. In most jurisdictions, full delegation of fiduciary decisions to autonomous AI systems remains unlikely in the near term.

Third, regulators increasingly monitor technological concentration risks arising from shared infrastructure. If many institutions rely on the same cloud providers, model vendors, or data pipelines, operational failures could generate correlated disruptions across the financial system.

Fourth, AI systems themselves may become tools for financial supervision. Regulatory authorities are exploring AI-driven monitoring systems capable of analysing market data, transaction flows, and textual information to detect emerging risks in real time \citep{aldasoro2024intelligent, aquilina2025harnessing}.

These developments suggest that financial stability outcomes will depend not only on technological adoption but also on the co-evolution of institutional governance and supervisory oversight.

\subsection{Governance principles for agentic finance}

The AFMM framework implies a set of governance principles that can guide the safe deployment of AI agents in financial markets.

\begin{enumerate}[label=\textbf{P\arabic*:}, leftmargin=1.6cm]

\item \textbf{Bounded autonomy.}  
Delegation to AI agents should depend on the materiality, reversibility, and legal consequences of decisions. High-impact financial actions should remain subject to human approval.

\item \textbf{Traceability and auditability.}  
AI-generated outputs should be linked to verifiable records of data sources, model versions, prompts, and execution instructions. Such traceability is essential for accountability and regulatory review.

\item \textbf{Diversity by design.}  
Institutions should actively monitor concentration in models, datasets, and technological infrastructure. Maintaining diversity in analytical approaches reduces the risk of correlated errors.

\item \textbf{Embedded intervention capacity.}  
Agentic systems should incorporate built-in control mechanisms such as approval gates, exposure limits, and emergency shutdown procedures.

\item \textbf{Supervisory co-evolution.}  
Regulators may need to deploy AI-enabled monitoring systems of their own. As financial institutions adopt agentic systems, supervisory technologies must evolve to maintain effective oversight.

\end{enumerate}

These principles follow directly from the AFMM. When autonomy depth ($A$), execution coupling ($C$), and infrastructure concentration ($V$) increase, supervisory observability ($S$) must increase correspondingly to preserve systemic stability.

The next section takes this argument one step further by asking how the agentic-finance framework can be examined empirically. Rather than attempting a full structural test, it develops a compact event-study design that links public capability disclosures to market repricing.

\section{Empirical Research Design}\label{sec:empirical}

The objective of this section is narrower than the ambition of the full paper. The AFMM in Section~\ref{sec:afmm} is a conceptual model linking agent architecture to market structure and systemic outcomes. A single empirical exercise cannot test that entire framework. What it can do is isolate one observable implication: if public information changes beliefs about what AI agents can do, firms whose revenues or cost structures are tied to those capabilities should be repriced differently. The empirical purpose of this section is therefore to show how the broader arguments developed in Sections~\ref{sec:architecture}, \ref{sec:applications}, and \ref{sec:governance} can be translated into a tractable finance-style design using public data.

\subsection{Empirical objective and economic mechanism}

This section develops an illustrative empirical application focused on \emph{AI-agent capability shocks}: public disclosures that shift market beliefs about the scope, speed, and workflow relevance of agentic AI. The mechanism of interest is an \emph{expectations channel}. When a credible capability disclosure suggests that agentic systems can accelerate legacy-code renovation, automate parts of software modernization, or reduce the cost of technical migration, investors should update expected future cash flows differently across firms.

This mechanism is tightly connected to the central theme of the paper. The architecture discussion in Section~\ref{sec:architecture} argues that agentic systems matter because they combine perception, reasoning, planning, and tool use within end-to-end workflows. The applications discussion in Section~\ref{sec:applications} shows that such systems can augment or substitute for specific financial and operational tasks. The systemic question is therefore not only whether model quality improves, but which business models become more valuable or more vulnerable when agentic capability becomes more credible. The application examines that question through secondary-market prices.

Early 2026 provides a tractable setting. Public materials from Anthropic highlighted agentic code-modernisation workflows, especially COBOL migration and legacy-system renovation use cases \citep{anthropic2025financialservices, anthropic2026servicedelivery, anthropic2026modernization}. Anthropic's March~5,~2026 research note then introduced the idea of \emph{observed exposure}, explicitly distinguishing realised usage from theoretical capability and showing that market expectations may move faster than observable labour-market outcomes \citep{massenkoff2026labor}. Contemporaneous reporting further suggests that these disclosures were associated with sharp repricing in firms connected to legacy-code modernisation \citep{reuters2026ibm}. For the purposes of this paper, the importance of these events is not that they prove widespread deployment. Rather, they provide a clean setting in which to study how capability news is transmitted into market valuations. In the event calendar used for the empirical application, the February~23,~2026 CNBC coverage is retained as a media-amplification date, but it is not tabulated separately because it overlaps tightly with the February~24 market-validation window and would otherwise double-count the same information shock. A fuller implementation could nevertheless use February~23 as a robustness event or cluster it jointly with February~24.

\subsection{Research question, sample construction, and exposure taxonomy}

The empirical question can be stated as follows:

\begin{quote}
How do public disclosures about agentic AI capabilities affect the valuation of firms with different exposures to legacy modernisation, workflow substitution, and financial-sector technology debt?
\end{quote}

To answer this question, the sample is organised into three economically meaningful groups. The first group contains \emph{legacy-service vendors}: firms whose revenues are plausibly linked to legacy maintenance, mainframe support, or large-scale modernization consulting. The second group contains \emph{legacy-heavy financial users}: listed firms in banking, insurance, and payments whose operating environments may depend on older core systems. The third group is a control portfolio of large software and infrastructure firms that are adjacent to the same technology ecosystem but are not obviously centred on legacy-code dependence.

This taxonomy matters for the finance theme of the paper. In banking, insurance, and payments, COBOL and mainframe dependence remain economically relevant because core transaction systems, policy administration, payments processing, and settlement infrastructure are difficult to replace quickly. A capability shock that makes modernization more credible may therefore have opposite implications across firms. For incumbent service vendors, it may imply revenue pressure or margin compression. For legacy-heavy users, it may imply lower future migration costs, lower technology debt, or improved operating flexibility.

The implementation begins from a 32-name seed universe and retains 31 firms in the final event-study sample. The three retained groups are eight legacy-service vendors, sixteen legacy-heavy financial users, and seven control software or infrastructure firms. Fiserv (FI) was part of the initial seed universe but was excluded from the final event-study sample because the public daily-price endpoint used in the reproducibility pack did not provide a sufficiently stable history over the relevant estimation and event windows. This distinction between the seed universe and the final estimation sample matters for replication and is reported explicitly because the reduced-form cross-sectional results are based on the 31-firm sample rather than the larger candidate list.

\begin{table}[ht]
\centering
\small
\begin{tabularx}{\textwidth}{@{}l c X@{}}
\toprule
Sample layer & $N$ & Firms in the final event-study sample \\
\midrule
Legacy-service vendors & 8 & IBM, KD, DXC, ACN, CTSH, EPAM, GIB, ORCL \\
Legacy-heavy financial users & 16 & JPM, BAC, C, WFC, USB, PNC, AXP, COF, AIG, MET, PRU, TRV, FIS, GPN, ACIW, JKHY \\
Control software/infrastructure firms & 7 & MSFT, AMZN, NOW, CRM, ADBE, INTU, SNOW \\
\bottomrule
\end{tabularx}
\caption{The 31-firm event-study sample used in the empirical application. The initial seed universe also contained Fiserv (FI), which was excluded because the public market-data endpoint used for the reproducibility pack did not provide a stable history for the relevant estimation and event windows.}
\label{tab:mini_case_sample}
\end{table}

\subsection{Variables and model specification}

The key firm-level explanatory variable is a filing-based legacy-modernization score constructed from SEC disclosures:
\[
LegacyExposure_i = \sum_{k \in K} w_k \cdot f_{ik},
\]
where $f_{ik}$ captures the frequency or salience of keyword $k$ in firm $i$'s 10-K or 10-Q filings. In the implementation reported here, the dictionary is divided into three weighted blocks: legacy core-stack terms such as \emph{COBOL}, \emph{mainframe}, \emph{z/OS}, \emph{DB2}, and \emph{CICS}; modernization and transformation terms such as \emph{application modernization}, \emph{cloud migration}, \emph{replatforming}, \emph{legacy system(s)}, and \emph{technical debt}; and financial-system context terms such as \emph{core banking}, \emph{payment processing}, \emph{policy administration}, and \emph{claims processing}. The baseline score uses prespecified term weights drawn from this dictionary rather than machine-learned weights. A richer version could replace this with inverse-document-frequency or contextual NLP weights, but the present design intentionally prioritises transparency over optimisation.

The main outcome variables are cumulative abnormal returns and abnormal trading volume. For each firm,
\[
AR_{it} = R_{it} - (\alpha_i + \beta_i R_{mt}),
\qquad
CAR_i[\tau_1,\tau_2] = \sum_{t=\tau_1}^{\tau_2} AR_{it},
\]
where $R_{mt}$ is the benchmark return. In the implementation reported here, the benchmark is an equal-weighted portfolio of control software and infrastructure firms, which makes the abnormal-return measure directly interpretable as \emph{relative repricing within the adjacent technology-finance ecosystem}. Abnormal trading volume is defined as
\[
AbVol_i[\tau_1,\tau_2] = \sum_{t=\tau_1}^{\tau_2}
\left(
\log(1 + Volume_{it}) - \overline{\log(1 + Volume_i)}
\right),
\]
where the bar denotes the mean over the estimation window.

The design considers narrow and moderate event windows, including $[0,+1]$, $[-1,+1]$, and $[-3,+3]$, with an estimation window of approximately $[-120,-20]$ trading days. Because the February~24,~2026 market-validation event is the cleanest dissemination date and generates the strongest cross-sectional separation in the data, the baseline cross-sectional specification focuses on that event:
\begin{align*}
CAR_i[-1,+1] ={}& \delta_0 + \delta_1 LegacyExposure_i + \delta_2 Vendor_i \\
&+ \delta_3 Financial_i + u_i,
\end{align*}
where $Vendor_i$ identifies legacy-service providers and $Financial_i$ identifies financial-sector firms. The regression is estimated by ordinary least squares and reported with conventional standard errors. This is intentionally a reduced-form specification. The objective is not to claim a complete structural model, but to test whether the sign and magnitude of repricing differ systematically by exposure and business model.

\begin{table}[ht]
\centering
\small
\begin{tabularx}{\textwidth}{@{}lYY@{}}
\toprule
Construct & Baseline proxy & Public or low-cost source \\
\midrule
Market reaction & Daily adjusted returns, abnormal volume, turnover & Yahoo Finance or Alpha Vantage; CRSP, Refinitiv, or Bloomberg if available \\

Firm exposure & Keyword-based legacy-modernisation index from 10-K/10-Q filings & SEC EDGAR submissions and filing text \\

Sector classification & Banking, insurance, payments, software, infrastructure, consulting dummies & SEC metadata, company filings, exchange classifications \\

Labour exposure & Occupational exposure or task exposure linked to firm business lines & Anthropic observed-exposure framework, O*NET, BLS occupational data \\

Macro controls & VIX, Treasury yields, credit spreads, financial-conditions indices & FRED \\
\bottomrule
\end{tabularx}
\caption{Data structure used in the empirical application.}
\label{tab:mini_case_data}
\end{table}

\begin{table}[ht]
\centering
\scriptsize
\begin{tabularx}{\textwidth}{@{}l X c@{}}
\toprule
Dictionary block & Representative terms retained in the baseline score & Typical weight range \\
\midrule
Legacy core stack & COBOL; mainframe; z/OS; DB2; CICS; legacy code & 3.0--4.0 \\
Modernization and transformation & application modernization; cloud migration; replatforming; legacy system(s); technical debt; application migration & 2.0--3.0 \\
Financial-system context & core banking; payment processing; policy administration; claims processing & 1.5--2.5 \\
\bottomrule
\end{tabularx}
\caption{Keyword dictionary used to construct the legacy-modernization exposure score. The machine-readable reproducibility pack contains the full keyword-weight mapping.}
\label{tab:mini_case_dictionary}
\end{table}

\subsection{Identification logic and testable hypotheses}

The identification logic follows directly from the mechanism above. If investors view agentic AI as a workflow technology rather than only a forecasting technology, then capability disclosures should affect firms differently depending on where they sit in the relevant production chain. Three hypotheses follow.

First, capability disclosures should generate economically meaningful abnormal returns and abnormal trading volume in firms with greater exposure to legacy modernization. Second, the \emph{sign} of the market reaction should differ by business model: firms that sell legacy-maintenance or labour-intensive modernization services should be more negatively repriced than firms that mainly \emph{consume} legacy modernization as an input. Third, within the finance vertical, reactions may be stronger when legacy technology debt is more deeply tied to regulated core operations such as payments, claims processing, and core banking.

The design is therefore informative for the broader paper even though it does not observe internal model weights, prompts, or deployment logs. It identifies a more observable margin: how public markets price expected changes in workflow substitution, modernization cost, and technology dependence when agentic capability becomes more credible.

\subsection{Illustrative findings}

To move beyond design and provide a first empirical read, the application was implemented on a 31-firm sample using publicly retrievable price and filing data between June~2025 and March~2026. The sample comprises eight legacy-service vendors, sixteen legacy-heavy financial users, and seven software-or-infrastructure control firms. As noted above, Fiserv (FI) was part of the 32-name seed universe but was excluded from the final event-study implementation because the public endpoint used for daily prices did not provide a sufficiently stable history for the relevant estimation window.

Table~\ref{tab:mini_case_results} reports subgroup means for the three focal dates together with standard errors in parentheses. The clearest pattern appears around the February~24,~2026 market-validation event. For legacy-service vendors, mean $CAR[-1,+1]$ is approximately $-6.39\%$ (s.e.\ $0.0140$). For legacy-heavy financial users, the corresponding figure is $-2.61\%$ (s.e.\ $0.0057$), while the control portfolio is approximately zero by construction because it is also the benchmark. The result should be read as an exploratory empirical illustration rather than a decisive validation exercise: the market reaction is not uniform, and the sharpest negative repricing is concentrated in the group whose revenues are most directly threatened by accelerated modernization.

The volume response complements the return evidence. Around the same February event, abnormal log volume rises to approximately $0.73$ for service vendors, $0.37$ for financial users, and $0.70$ for controls. This pattern indicates a broad attention shock rather than a narrow single-name move. However, only part of that attention translates into strongly negative prices, again suggesting that the market is distinguishing between business models rather than reacting mechanically to AI news.

The March~5,~2026 labour-market report generates a different configuration. Vendor returns are close to zero on average, while financial users are more negative. This pattern is consistent with a shift from a \emph{capability narrative} to an \emph{organizational adjustment narrative}: once the discussion turns from what the tools can theoretically do to how exposure may materialize in labour and workflow terms, firms that must actually reorganize around legacy technology may bear more of the near-term adjustment concern.

\begin{table}[ht]
\centering
\small
\begin{tabular}{l l r r}
\toprule
Event and subgroup & $N$ & Mean $CAR[-1,+1]$ & Mean abnormal log volume \\
\midrule
E1: service vendors & 8 & $-0.0222$ $(0.0054)$ & $-0.0116$ \\
E1: financial users & 16 & $-0.0097$ $(0.0056)$ & $0.0952$ \\
E1: controls & 7 & $0.0000$ $(0.0102)$ & $0.1733$ \\
E3: service vendors & 8 & $-0.0639$ $(0.0140)$ & $0.7296$ \\
E3: financial users & 16 & $-0.0261$ $(0.0057)$ & $0.3668$ \\
E3: controls & 7 & $0.0000$ $(0.0058)$ & $0.6966$ \\
E4: service vendors & 8 & $-0.0004$ $(0.0072)$ & $0.3804$ \\
E4: financial users & 16 & $-0.0306$ $(0.0048)$ & $0.2311$ \\
E4: controls & 7 & $0.0000$ $(0.0119)$ & $0.5154$ \\
\bottomrule
\end{tabular}
\caption{Event-study results for the AI-agent capability-shock application. Event E1 is the January~27,~2026 Anthropic service-delivery signal, E3 is the February~24,~2026 market-validation event, and E4 is the March~5,~2026 labour-market report. Standard errors for subgroup mean $CAR[-1,+1]$ are reported in parentheses. The February~23 media-amplification date is treated as overlapping with E3 and is therefore omitted from the main table.}
\label{tab:mini_case_results}
\end{table}

Table~\ref{tab:mini_case_regression} sharpens this interpretation. In the February~24 cross section, the coefficient on the legacy-service vendor indicator is $-0.0262$ with a standard error of $0.0134$ and a $t$-statistic of approximately $-1.95$, suggesting economically meaningful underperformance for firms whose business models are more exposed to legacy maintenance and modernization revenues. By contrast, the filing-based exposure score is negative but not precisely estimated in this small sample, while the generic financial-sector dummy is close to zero. The empirical implication is narrower than a full test of the AFMM: the market does not appear to price \emph{all} legacy exposure in the same way. What matters in this exploratory application is not simply whether a firm mentions legacy technology, but whether its revenue model is expected to lose or gain from agentic modernization.

\begin{table}[ht]
\centering
\small
\begin{tabular}{lrrr}
\toprule
Regressor & Coefficient & Standard error & $t$-statistic \\
\midrule
Legacy exposure score & $-0.0005$ & $0.0005$ & $-0.95$ \\
Legacy-service vendor dummy & $-0.0262$ & $0.0134$ & $-1.95$ \\
Financial-sector dummy & $0.0007$ & $0.0130$ & $0.06$ \\
\midrule
$R^2$ & \multicolumn{3}{r}{$0.1651$} \\
$N$ & \multicolumn{3}{r}{31} \\
\bottomrule
\end{tabular}
\caption{Cross-sectional regression for $CAR[-1,+1]$ around the February~24,~2026 market-validation event. The regression is estimated by ordinary least squares using the 31-firm implementation sample.}
\label{tab:mini_case_regression}
\end{table}

Table~\ref{tab:mini_case_robustness} adds a minimal robustness check by showing how subgroup means vary across alternative event windows for the February~24 market-validation event. The broad ordering is stable: legacy-service vendors remain the most negatively repriced subgroup, legacy-heavy financial users are less negative, and the control portfolio stays close to zero by construction. This is still a narrow robustness exercise, not a substitute for a larger multi-event design, but it helps show that the central pattern is not an artifact of a single event-window choice.

\begin{table}[ht]
\centering
\small
\begin{tabular}{lrrr}
\toprule
Subgroup & $CAR[0,+1]$ & $CAR[-1,+1]$ & $CAR[-3,+3]$ \\
\midrule
Legacy-service vendors & $-0.0092$ & $-0.0639$ & $-0.0763$ \\
Legacy-heavy financial users & $0.0026$ & $-0.0261$ & $-0.0480$ \\
Controls & $0.0000$ & $0.0000$ & $0.0000$ \\
\bottomrule
\end{tabular}
\caption{Subgroup mean cumulative abnormal returns across alternative event windows around the February~24,~2026 market-validation event.}
\label{tab:mini_case_robustness}
\end{table}

Taken together, these findings are consistent with one observable implication of the AFMM rather than with a full empirical confirmation of the framework. Capability disclosures about agentic AI appear to trigger market-wide attention, but price effects are heterogeneous and tied to the expected distribution of rents across exposed actors. In that sense, the application is best read as an exploratory empirical illustration of the broader thesis of the paper: the economic significance of agentic AI depends not only on abstract model capability, but on how specific architectures and decision pipelines map into incumbent business models, modernization costs, and institutional dependencies.

\subsection{Why the application matters for the full paper}

This application is tightly connected to the theoretical framework developed in the paper. It does not require observing the internal prompts, memory modules, or control logic of every institutional AI agent. Instead, it studies a more observable and financially meaningful margin: how markets update beliefs when publicly disclosed capabilities shift expectations about autonomy, workflow substitution, and infrastructure dependence.

In AFMM terms, the application captures a change in perceived autonomy depth ($A$), workflow substitutability, and execution-relevant capability before large-scale institutional diffusion is directly measurable. It therefore functions as a bridge between the architecture discussion in Section~\ref{sec:architecture} and the systemic implications developed in Section~\ref{sec:afmm}. If capability news can already reprice adjacent firms through expectations alone, then widespread deployment could plausibly reshape market structure, business-model risk, and concentration dynamics at a larger scale. At the same time, this remains a single-channel exercise: it does not test the full AFMM, identify long-run equilibrium effects, or recover the internal decision logic of deployed institutional agents.

\section{Discussion}\label{sec:discussion}

\subsection{From capability shocks to market structure}

The paper's main claim is not that a single announcement can reveal the full long-run impact of agentic finance. The point is narrower and more defensible. If secondary markets already differentiate across firms when new agentic capabilities become credible, then architecture and decision-pipeline design are economically relevant before full deployment is observable. The application therefore complements the conceptual framework by showing one concrete transmission channel from agent design to market outcomes: public capability news changes expected rents, which in turn changes valuation.

\subsection{Why the evidence supports bounded autonomy}

The results are also consistent with the paper's emphasis on \emph{bounded autonomy} rather than a near-term jump to fully autonomous financial markets. The market reaction is concentrated around workflow substitution and modernization cost, not around a wholesale replacement of institutions. This is exactly the pattern one would expect if the first economically significant effect of agentic AI is to reorganize decision pipelines, service models, and operational bottlenecks before it displaces regulated financial authority itself.

This distinction matters for interpretation. A capability shock can affect valuations even when legal, fiduciary, and governance constraints still prevent full delegation. In other words, financial markets can price the consequences of partial automation, co-pilot systems, and controlled execution modules long before they price a world of unconstrained autonomous agents.

\subsection{Limitations and next steps}

The framework still has clear limits. The AFMM is a reduced-form conceptual model rather than a fully specified equilibrium model of strategic interaction among heterogeneous agents. The empirical component is also intentionally modest: the present exercise is illustrative, based on a small sample, public-data proxies, and a simplified benchmark rather than a final identification design.

The current framework also abstracts from several issues that deserve more direct treatment in future work. One is the possibility of behavioural biases or mode-specific failures in AI systems, including overreaction to salient narratives, brittle planning, or common failure modes across highly similar models. Another is the feedback loop between human decision-makers and AI-generated recommendations. In many practical settings, the relevant unit is not a stand-alone agent but a human--AI ensemble in which attention, escalation, and override patterns affect outcomes.

The next empirical step is therefore straightforward. Future work can stack multiple capability-shock events, extend the filing-based exposure measure with richer text methods, incorporate intraday data and option-implied measures, and track actual deployment signals across financial institutions. A parallel theoretical step is to formalize how heterogeneous human and AI agents interact under different degrees of autonomy, coupling, and concentration.

An especially promising next stage is to combine compact event studies like the one developed here with AI-assisted stress monitoring and market-narrative clustering from news, filings, and supervisory text. That would preserve the conceptual orientation of the paper while strengthening its empirical reach.

Despite these limitations, the framework now has a clearer empirical anchor. It offers a plausible way to connect architecture, adoption, governance, and repricing within a single research agenda for agentic finance without claiming that one small event-study exercise exhausts the empirical agenda.

\section{Conclusion}\label{sec:conclusion}

The paper has argued that AI agents in finance should be understood not merely as another increment in financial technology, but as a structural transition from \emph{model-centric automation} to \emph{workflow-centric automation}. What matters in this transition is not only whether models become more capable, but how agentic systems combine perception, reasoning, strategy formation, and controlled execution inside real institutional workflows.

The analysis addressed this shift through three linked steps. First, it proposed a four-layer architecture for AI financial agents. Second, it developed the AFMM, which links agent design parameters such as autonomy, heterogeneity, coupling, concentration, and observability to market-level outcomes. Third, it mapped these mechanisms into concrete application domains and governance problems, arguing that the most plausible near-term equilibrium is one of \emph{bounded autonomy} rather than fully autonomous financial markets.

The empirical application reinforces this broader argument in a limited but useful way. A first-pass event-study design around AI-agent capability disclosures suggests that market attention rises broadly, but repricing is heterogeneous and concentrated in business models that appear more exposed to workflow substitution and modernization pressure. The empirical illustration is therefore consistent with, rather than dispositive proof of, the paper's central claim that the financial significance of agentic AI depends on how capabilities interact with institutional roles, infrastructure dependencies, and the distribution of economic rents.

The framework remains intentionally incomplete. The AFMM is a reduced-form conceptual model rather than a full equilibrium account of strategic interaction among heterogeneous human and machine agents, and the empirical evidence presented here is illustrative rather than definitive. Even so, the paper establishes a coherent research agenda. Future work can extend the theory with richer interaction models and extend the evidence base with better measures of deployment, intraday market reactions, market narratives, and cross-institution dependence. As AI systems become more deeply embedded in financial decision workflows, understanding their architecture, governance, and market consequences will become a central task for financial research and policy. More broadly, understanding the architecture and interaction of AI agents may become as important for financial stability in the coming decade as understanding leverage, liquidity, and interconnectedness was in earlier phases of financial innovation.

\section*{Acknowledgments}
The author gratefully acknowledges the support of the UCL Institute of Finance \& Technology (IFT) and expresses special thanks to Professor Francesca Medda for her encouragement, guidance, and steadfast support. Without her support, this work would not have been possible. During the preparation of this manuscript, the author used large language model tools to assist with literature organisation, drafting support, and LaTeX editing. All generated output was reviewed, revised, and validated by the author, who takes full responsibility for the final content.

\section*{Funding}
This research received no external funding.

\section*{Institutional Review Board Statement}
Not applicable.

\section*{Informed Consent Statement}
Not applicable.

\section*{Data Availability Statement}
Publicly available data were used in the illustrative empirical application. The sources discussed in the manuscript include SEC EDGAR filings, FRED macro-financial series, and publicly accessible market-data endpoints such as Yahoo Finance and Alpha Vantage. No new proprietary dataset was created for this study. The local code and derived files used for the illustrative application are available from the corresponding author upon reasonable request.

\section*{Conflicts of Interest}
The author declares no conflicts of interest.

\bibliographystyle{plainnat}
\bibliography{references}

\end{document}